\documentclass{article}
\usepackage[T1]{fontenc}
\usepackage[utf8]{inputenc}
\usepackage{units}
\usepackage{amstext}
\usepackage[numbers]{natbib}
\usepackage{microtype}
\usepackage[unicode=true,
 bookmarks=false,
 breaklinks=false,pdfborder={0 0 1},backref=section,colorlinks=false]
 {hyperref}

\makeatletter

\usepackage[final]{nips_2018}
\usepackage{subfigure} 
\usepackage{graphicx} % more modern
\usepackage{colortbl}
\usepackage{url}% simple URL typesetting
\usepackage{amsfonts}% blackboard math symbols
\usepackage{nicefrac}% compact symbols for 1/2, etc.
\title{Formatting instructions for NIPS 2018}

\author{
  Yordan P. Raykov${^{1,\dagger}}$, Luc J.W. Evers${^{2,\dagger}}$, Reham Badawy${^1}$, Marjan J. Faber${^2}$,\\
  \textbf{Bastiaan R. Bloem${^2}$, Kasper Claes${^3}$, Max A. Little$^{1,4}$}\\
  $^1$Department of Mathematics, Aston University, United Kingdom \\
  $^2$Department of Neurology, Radboud university medical center, Netherlands\\
  $^3$UCB Pharma, Belgium \\
  $^4$Media Lab, Massachussetts Institute of Technology, United States \\
 }
\makeatother

\begin{document}
\title{Probabilistic modelling of gait for remote passive monitoring applications}
\maketitle
\begin{abstract}
Passive and non-obtrusive health monitoring using wearables can potentially
bring new insights into the user's health status throughout the day
and may support clinical diagnosis and treatment. However, identifying
segments of free-living data that sufficiently reflect the user's
health is challenging. In this work we have studied the problem of
modelling real-life gait which is a very indicative behaviour for
multiple movement disorders including Parkinson's disease (PD). We
have developed a probabilistic framework for unsupervised analysis
of the gait, clustering it into different types, which can be used
to evaluate gait abnormalities occurring in daily life. Using a unique
dataset which contains sensor and video recordings of people with
and without PD in their own living environment, we show that our model
driven approach achieves high accuracy gait detection and can capture
clinical improvement after medication intake.
\end{abstract}

\section{Introduction}

Sensors embedded in smartphones and wearables can provide valuable
information about the health status of patients with various movement
disorders \citep{Lipsmeier2018}. Most research has focused on analyzing
standardized assessments in controlled environments \citep{arora2014high}.
When deployed in the patient's own home environment, wearable sensors
could be used to passively record real-life symptom fluctuations \citep{De2016}.
This could provide physicians and patients with additional information
about the progression of the disease and the efficacy of treatment,
even without the need for hospital visits. The modelling challenge
this introduces is how to deal with the enormous variation in real-life
signals, which is not only influenced by the patient's condition,
but also by the large variation in daily life behaviours. One approach
is to analyze behaviour that is common across patients and likely
to reflect the status of the patient's condition. An important example
of such behaviour is gait, a highly prevalent and stereotypic behaviour,
which is known to be indicative of several neurological conditions,
such as Parkinson's disease (PD). 

There is a plethora of prior work focused on analysis of gait using
accelerometers and gyroscopes \citep{brajdic2013,Avci2010,He2008,Iso2006}.
Most algorithms consist of two separate steps, i.e. gait detection
and for quantification of the gait pattern. The detection phase normally
aims to segment out all activities which involve walking while the
further analysis of these segments is done to discover periods of
gait and gait properties indicative of user's health. Gait detection
is often done using supervised learning or thresholding techniques,
applied to simple features extracted from a moving window of fixed
width. Commonly used features include standard deviation \citep{Dijkstra2010detection,Godfrey2016,Del2016free_living};
Fourier coefficients \citep{Karantonis2006,Moore2011,mannini2013}
and normalized autocorrelation \citep{Rai2012}. Hand tuned thresholds
can be avoided by training classification methods such as Support
Vector Machines (SVMs) \citep{Sama2017}, Naive Bayes or 2-component
Gaussian Hidden Markov Models (HMMs) \citep{Mannini2011activity,Nickel2011benchmarking}.
From the identified gait, properties such as step time and step length
can be estimated to describe the quality of the user's gait. These
methods, often developed for specific body locations, rely on identifying
specific peaks that correspond to the initial and final contact of
the gait cycle \citep{Randell2003,Kim2004,Mannini2011gait}. Less
location-specific are spectral methods that look at the leading harmonics
\citep{brajdic2013} and autocorrelation methods that measure autocorrelation
at certain lags \citep{Rai2012}. 

Most of the methods for both detection and analysis of gait are validated
on labeled datasets that only include a restricted set of activities
performed in controlled environments. Because of this and their lack
of rigorously dealing with uncertainty, such methods have hard to
analyze behaviour when applied to data collected in daily life, where
an arbitrary large number of different activities can be recorded.
Whereas some empirical prove of accuracy might be sufficient in consumer
applications, medical applications require highly interpretable algorithms
that can generate reliable and verifiable insights in daily life gait
patterns. 

In this work our objective is to jointly detect and analyze gait patterns
combining previously proposed ideas into a robust model based framework
for analysis of real-life gait data. Our approach extends existing
HMM techniques by adopting a nonparametric framework that allows us
to infer different gait and non-gait states from the data in an unsupervised
way. Also, we propose using an autoregressive observation model to
describe each state. We demonstrate our approach on a new unique reference
dataset consisting of sensor data from various wearables and concurrent
video annotations, collected in the home environment of people with
and without Parkinson's disease (PD). 

\section{Probabilistic modelling of gait}

In this section we layout the minimalistic assumptions that we embed
in our probabilistic model for analysis of gait data from wearable
accelerometer sensors. We do not extract many types of correlated
features that are hard to interpret but we attempt to model a minimal
amount of features (the acceleration amplitude) and account for some
of the common confounding factors occurring in data recorded by inertial
measurement units. This more direct model-driven approach to gait
analysis is less reliant on labeled data which is an essential requirement
for future tools for passive monitoring.

\textit{Pre-processing}: Accelerometer data collected from different
smartphones is sampled at non-uniform rates varying from approximately
30 to 200Hz. Therefore, data is interpolated to a uniform rate and
further downsampled (using standard anti-aliasing moving average filter)
to 30Hz, which is sufficient to measure human gait \citep{angeloni1994frequency}.
Accelerometers are heavily confounded by changes in the orientation
of the device irrelevant to the user's gait. In order to remove this
undesired effect of gravity, we use a L1-trend filter \citep{Kim2009,badawy2018}
to estimate a piecewise linear vector which is then subtracted from
the accelerometer data. Finally, we model the amplitude of the remaining
dynamic component which we can denote as a 1-D series $x_{1},\dots,x_{T}$. 

\textit{Capturing the rhythm of gait:} Natural human gait is highly
repetitive as it consists of multiple consecutive gait cycles. As
a result, the accelerometer signal during gait is generally highly
periodic. However, many other periodic and non-periodic movements
occur in free-living human behaviour. To model free-living gait, we
need both high frequency resolution to capture underlying patterns
of gait and precise time localization of when gait starts or stops.
This is hard to achieve in the Fourier domain and we usually sacrifice
time localization accuracy. But if we apply a sequential switching
state space mechanism, we can adaptively infer the start and end of
the stationary intervals associated with the gait behaviour. Here
we model the spectrum of the gait using autoregressive processes (AR).
An order $r$ AR model is a random process which describes a sequence
$x_{t}$ (for $t=r+1,\dots,T$) as a linear combination of previous
values in the sequence and a stochastic term: $x_{t}=\sum_{j=1}^{r}A_{j}x_{t-j}+e_{t}$
with $e_{t}\sim\mathcal{N}\left(0,\sigma^{2}\right)$ where $A_{1},\dots,A_{r}$
are the AR coefficients and $e_{t}$ is a zero mean, Gaussian i.i.d.
sequence. An important property of AR models is that we can express
its power spectral density as a function of its coefficients: $S\left(f\right)=\frac{\sigma^{2}}{\left|1-\sum_{j=1}^{r}A_{j}\exp\left(-i2\pi fj\right)\right|^{2}}$
where $f\in\left[-\pi,\pi\right]$ is the frequency variable with
$i$ here denoting the imaginary unit. This means that the order of
the AR model directly determines the number of ``spikes'' in its
spectral density. We assume that with a high enough order $r$, an
AR model is a good summary of a particular type of gait \citep{Valenza2014}.
Next we need a systematic way to infer when behaviours other than
gait occur or when a different type of gait is being observed. To
do this in sufficiently flexible manner, we couple the AR component
with a nonparametric hidden Markov model \citep{Beal2002,Teh2004}
(more specifically: direct assignment HDP-HMM\citep{Teh2004}) to
obtain a nonparametric switching AR process, which was first introduced
in \citet{Fox2009}. In nonparametric switching AR processes we assume
that the data is an inhomogeneous stochastic process and multiple
different AR models are required to represent the dynamic structure
of the series, i.e.:
\begin{equation}
x_{t}=\sum_{j=1}^{r}A_{j}^{\left(z_{t}\right)}x_{t-j}+e_{t}^{\left(z_{t}\right)}\qquad e_{t}^{\left(z_{t}\right)}\sim\mathcal{N}\left(0,\left(\sigma^{\left(z_{t}\right)}\right)^{2}\right)
\end{equation}
where $z_{t}\in\left\{ 1,\dots,K^{+}\right\} $ indicates the AR model
associated with point $t$ and $K^{+}$ varies. The latent variables
$z_{1},\dots,z_{T}$ describing the switching process are modelled
with a Markov chain. This Markov chain is parametrized with a transition
matrix, the weights of which are modelled with a hierarchical Dirichlet
process (HDP). This allows the number of represented states $K^{+}$
to be inferred from the data where $K^{+}\ll T$ allows us to cluster
together data which is likely to be modelled with the same AR coefficients. 

\textit{Sparseness of the representation}: To finish our model, we
specify a prior over the remaining unknown parameters, i.e. the AR
coefficients and the process error. Since these can be seen as the
parameters of a linear regression problem, the conjugate prior over
$\left\{ A,\Sigma\right\} $ is a matrix Normal inverse Wishart prior
which for univariate $x$ collapses down to a multivariate Normal
inverse Gamma prior: $\left\{ A,\Sigma\right\} \sim\mathcal{N}\left(A\left|\mu_{0},\sigma I_{r}\right.\right)\text{Inv-Gamma}\left(\sigma\left|\nu,\theta\right.\right)$.
Despite of the numerical convenience of this prior, better control
over the sparseness of the posterior coefficients can be obtained
if we instead use a non-conjugate independent $0$-mean Gaussian prior\footnote{One can further assume $\sigma$ varies across coefficients in order
to place more importance on inferring patterns in the data that occur
with specific lags} over each coefficient, $A_{j}\sim\mathcal{N}\left(0,\sigma\right)$
for $j=1,\dots,r$. In this way we can model the AR coefficients and
error separately; we define a separate inverse Gamma prior over the
process error: $\sigma\sim\text{Inv-Gamma\ensuremath{\left(\nu,\theta\right)}}$
(this prior is referred to as automatic relevance determination prior
in \citep{Fox2009thesis}). The effect of the two priors is compared
visually on the same gait data in Appendix \ref{sec:Sparse-vs-non-sparse}. 

\begin{figure}[htbp]  
\centering
\includegraphics[width=0.75\columnwidth]{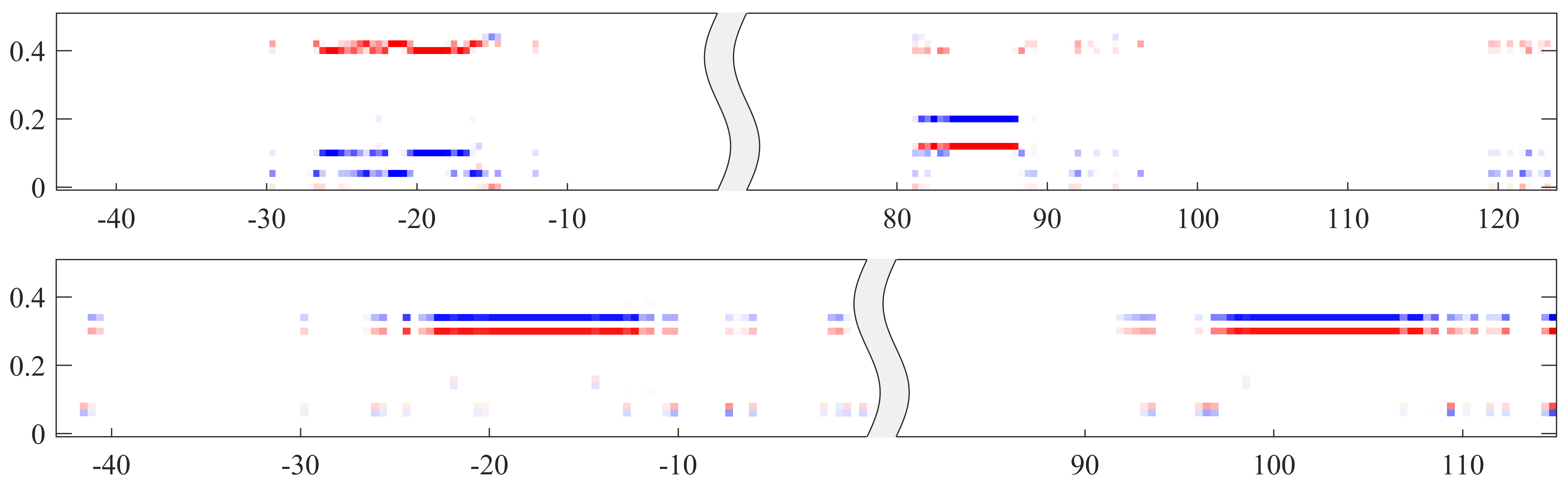}

\caption{Estimated features $c$ (red) and $a+d$ (blue) for a responsive (top) and a non-responsive (bottom) PD patient. $x$-axis is time relative to medication intake; $y$-axis is feature values.} 
\label{fig.BeforeAfter}
\end{figure}

\section{Evaluation}

We apply the proposed framework for model based gait analysis to data
from the Parkinson@Home validation study (see Appendix \ref{sec:Parkinson's@-Home-validation}).
In brief, 24 participants with PD receiving dopaminergic medication
and 24 age-matched participants without PD were visited in their own
homes. The visits included an unscripted free-living part of approximately
1 hour, in which the assessors encouraged the participants to perform
usual activities. The PD group was instructed not to take their PD
medication before the visit, in order to perform the free-living part
both before and after medication intake. Participants wore various
light-weight sensors and the full visits were recorded on video. In
this work, we use accelerometer data from a smartphone worn in the
front trouser pocket, and video annotations indicating the presence
and impairment of gait.

By segmenting the accelerometer signal in an interpretable way, we
aim to group various types of gait patterns and extract useful information
about the gait. After the accelerometer data from the users' smartphones
have been appropriately pre-processed, we fit in an unsupervised way
a nonparametric switching AR model of order $r=90$ to the amplitude
of the dynamic component. The nonparametric switching AR model adopts
an additional parameter supporting self-transitions (using 'sticky'
HDP \citep{fox2011}) and training is done using a novel streaming
slice sampling method. Similar results were obtained if using truncated
block sample from \citet{Fox2009}.

Our method can detect walking activities at least as well as other
best performing benchmark heuristics (selected from \citep{brajdic2013}).
Once we have trained the unsupervised switching AR on the data from
the home visits, there are several ways of using either some domain
knowledge or a small amount of supervised data to train a classifier
which predicts if data corresponds to gait related activity or not.
For example, here we train an interpretable multinomial Naive Bayes
classifier (which is linear in the log space) in which the input is
the posterior of $z_{t}$ for each point $t$ and the output is a
binary indicator denoting if a point is associated with a gait activity.
This approach relies on the assumption that most of the gait patterns
would be clustered in their own states. The hyperparameters for all
of the benchmarks are selected in-sample to maximize the balanced
accuracy and demonstrate highest possible discrimination accuracy
that these techniques can achieve on our new dataset. The standard
deviation thresholding (highest performing benchmark for gait detecion
in \citep{brajdic2013}) scored sensitivity of 0.79 and specificity
of 0.96; short-term Fourier transform gait detection scored sensitivity
of 0.85 and specificity of 0.94; normalized autocorrelation step counting
heuristics (NASC)\citep{Rai2012} scored sensitivity of 0.87 and specificity
of 0.94. The naive Bayes classifier just trained on the state indicators
associated with each point achieved out-of-sample sensitivity of 0.85
and specificity of 0.97 with standard deviation of 0.05 on both the
sensitivity and the specificity measures for different runs of the
switching AR model. 

In addition to identifying gait segments, the estimated AR parameters
associated with the data can be used to quantify some properties of
the observed gait. Given input sensor data of 30Hz sampling frequency,
the $r=90$ AR model estimates how well data lagged up to 3 seconds
predicts itself, i.e. the AR model estimates the periodicity of the
signal between bands of $0.33$Hz and $30$Hz. This means that information
such as the duration of the gait cycle, the variability of the gait
rhythm or its amplitude are already encoded in the inferred AR coefficients
and error. For example, assuming that our AR coefficients are fairly
sparse and the leading periodic component in the signal is due to
the repetitiveness of the full gait cycle, we can fit a Gaussian polynomial
$f\left(x\right)=a\exp\left(-\left(\frac{x-b}{c}\right)^{2}\right)+d$
to the AR coefficients to estimate both the expected time a gait cycle
takes and how much this time varies (see Appendix \ref{sec:Interpretation-of-AR}):
if the location $b$ is at the 40th coefficients, this would indicate
that the most likely cycle time is $\nicefrac{40}{30}$ seconds; $c$
would reflect how much this cycle time varies for a particular AR
state, and $a+d$ indicates the estimated value of $f$ at $b$ which
reflects the support at the peak cycle time $b$. 

We illustrate that the proposed features are highly related to medication-induced
changes in the gait pattern in Figure \ref{fig.BeforeAfter}, where
we plot the value of $c$ and $a+d$ for the fitted polynomials against
the time for two PD patients: one with clear clinically observable
changes in his gait pattern, and one with no observable changes. In
the first patient, we observe a consistent drop in the values of $c$,
a consistent increase in the values of $a+d$ and a mild increase
in $b$ (latter not plotted), whereas these changes are absent in
the non-responsive patient. Future work will address how these features
should be interpreted from a clinical perspective, and what are the
most important confounders (factors other than the severity of PD
symptoms) present in daily life that may influence the gait pattern
analysis.

\section{Conclusion}

Recent technological advances have made it possible to remotely and
continuously collect sensor data in daily life. Yet estimating features
that provide reliable insights in a person's health remains challenging.
For movement disorders such as PD, analyzing free-living gait is a
promising approach. Here we present a highly interpretable probabilistic
framework that can robustly detect gait and group different gait patterns
in PD patients corresponding with medication-induced changes. Advances
in algorithms to analyze passively collected sensor data will hopefully
provide novel insights in the real-life functioning of patients with
PD, which will benefit both clinical trials and individual patient
care.

$\textbf{\text{Acknowledgements:}}$ This work was supported by the
UCB Pharma, Michael J. Fox Foundation and Stichting Parkinson Fonds.
The authors extend their sincere gratitude to every individual who
participated in this study to generate the data used here.

\vfill{}

\bibliographystyle{abbrvnat2}

\newpage{}

\appendix

\section{Interpretation of AR coefficients when modelling gait\label{sec:Interpretation-of-AR}}

Visual example of the AR coefficients inferred for a segment of free
living gait. We fit a polynomial of selected form in red, where we
have chosen a Gaussian polynomial of the form $f\left(x\right)=a\exp\left(-\left(\frac{x-b}{c}\right)^{2}\right)+d$
and infer its best location according to the AR coefficient values.
Here we have done least squares polynomial fitting, but in the future
this step can also be done by Bayesian polynomial fitting. 

\begin{figure}[htbp]  
\centering
\includegraphics[width=0.5\columnwidth]{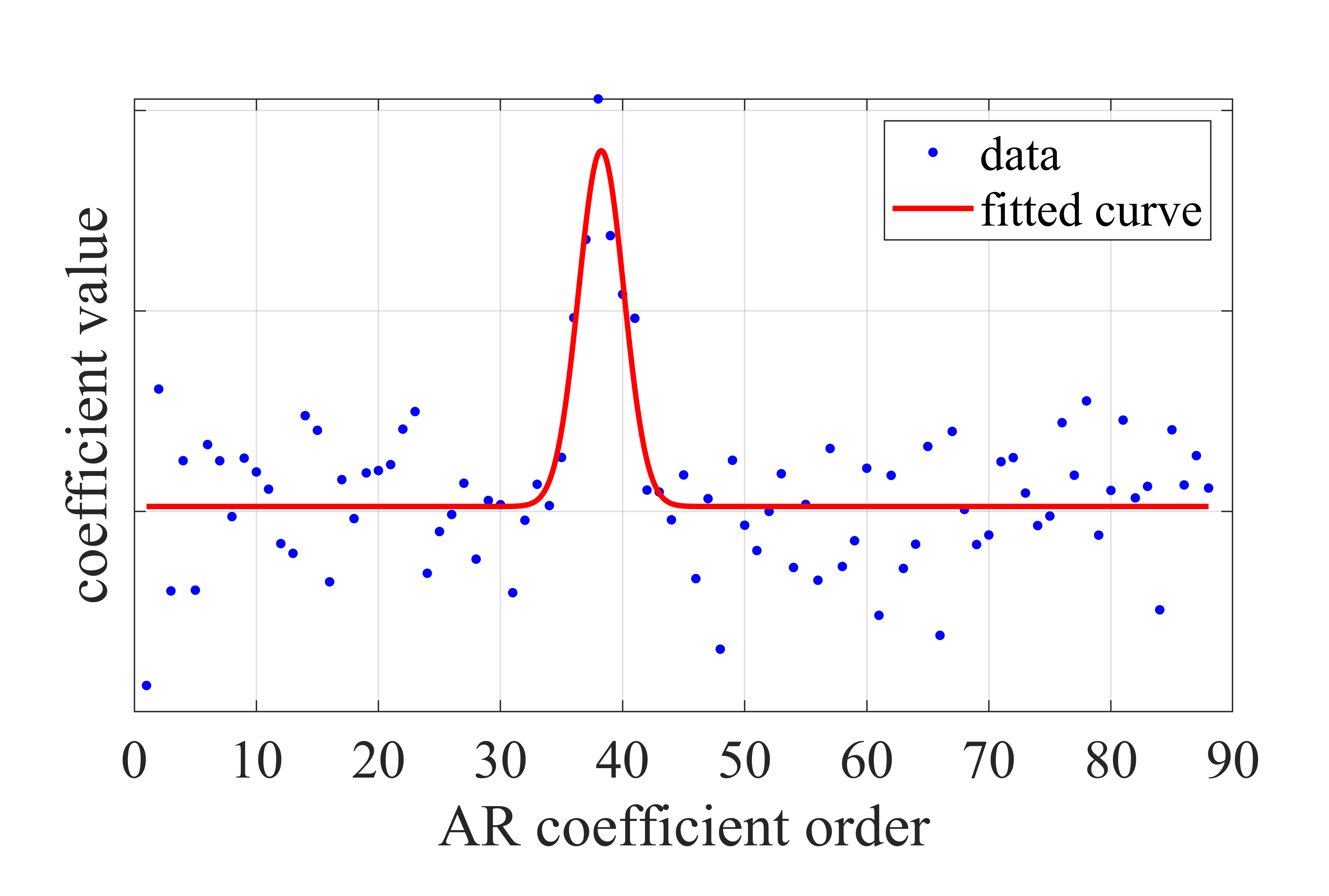}

\label{fig.ARcurveFitting}
\end{figure}

\section{Sparse vs non-sparse prior choice over the AR coefficients \label{sec:Sparse-vs-non-sparse}}

Visual comparison of the effect of non-conjugate sparse prior and
conjugate prior on the posterior of the AR coefficients estimated
by fitting an AR model on the same free living segment of gait using
Gibbs sampling. 

\begin{figure}[htbp]  
\centering
\includegraphics[width=0.5\columnwidth]{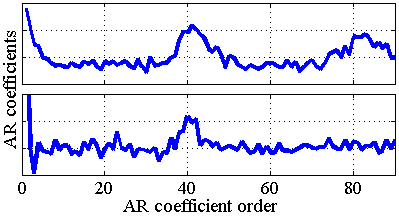}

\caption{The top graph displays the trained on gait data AR coefficients of an order 90 AR model assuming  sparse prior over the coefficients. The bottom graph displays the AR coefficients of an AR model of the same order trained on the same data but assuming conjugate prior over the coefficients. } 
\label{fig.ARcoefficientsComparisonPriors}
\end{figure}

\section{Parkinson's@ Home validation study - description\label{sec:Parkinson's@-Home-validation}}

Data used in this work was obtained from the Parkinson@Home validation
study, which was conducted at the Radboud University Medical Centre,
Nijmegen, The Netherlands.  In collaboration with the Michael J Fox
Foundation, a reference dataset will be made accessible to the research
community. The  study included 25 participants diagnosed with Parkinson’s
disease and 25 age-matched participants without Parksinson. Inclusion
criteria consisted of: (1) age \ensuremath{\ge}30 years and (2) in
possession of a smartphone running on Android 4.4 or higher. Additional
inclusion criteria for the Parkinson group were: (1) diagnosis of
Parkinson, (2) treated with levodopa and/or a dopamine agonist, (3)
experiencing motor fluctuations (MDS-UPDRS item 4.3 \ensuremath{\ge}
1), and (4) known with bradykinetic gait and/or freezing of gait (MDS-UPDRS
item 2.12 \ensuremath{\ge} 1 and/or item 2.13 \ensuremath{\ge} 1).
Candidates who received any advanced treatment were excluded. All
participants provided written informed consent and the study protocol
was approved by the local ethics committee (Commissie Mensgebonden
Onderzoek, Arnhem-Nijmegen; NL53034.091.15). Data was collected during
home visits in the morning, which  included  an unscripted free-living
part of approximately 1 hour, in which the assessors encouraged the
participants to perform usual activities. To make sure essential behaviors
were captured, such as longer gait cycles, assessors encouraged participants
to include these in their routines. The Parkinson group was asked
to skip their morning dose of dopaminergic medication before the visit.
The free-living part was repeated after medication intake. During
the full visit, participants wore various light-weight sensors on
both wrist, both shins, the lower back and in the front trouser pocket..
In this work, we used the accelerometer data obtained from the smartphone
worn in the trouser pocket. The complete visits were recorded on video,
which were annotated by a research assistant for the presence of walking
episodes, which were defined as as any activity including at least
5 consecutive steps, excluding stair climbing and running.

Currently, data was available for 48 participants because of data
quality issues in 2 participants: one smartphone sensor malfunctioning
and one problem with the video recordings.

\end{document}